# Accumulation-layer Surface Plasmons

SHIMA FARDAD,[1] E. ALEXANDER RAMOS,[2] ALESSANDRO SALANDRINO,[1,2,*]

[1]*Information and Telecommunication Technology Center, the University of Kansas, Lawrence KS 66045, USA*
[2]*Department of Electrical Engineering and Computer Science, the University of Kansas, Lawrence KS 66045, USA*
*Corresponding author: a.salandrino@ku.edu*

**A rigorous analytical study of the eigenmodes supported by a charge accumulation layer within a transparent conductive oxide (TCO) is presented. The new class of surface plasmons termed Accumulation-Layer Surface Plasmon (ASP) is introduced. Near resonance ASP are tightly bound and display a vast effective index tunability that could be of great practical interest. The suppression of ASP in the presence of epsilon-near zero regions is discussed.**

Transparent conductive oxides (TCOs) have attracted a great deal of interest in the past few years as alternative materials for plasmonics [1] in the near-infrared region. In contradistinction to noble metals, TCOs such as Indium Tin Oxide (ITO) display a vast tunability of their optical and electronic properties [2] via doping and electric bias. The possibility of actively switching between a low-loss dielectric regime and a high-absorption plasmonic regime has been exploited for the design and realization of ultra-compact electro-absorption modulators [3-7], as well as for the proposal of novel multimode modulator architectures [8].

At the heart of the applications outlined before is the electron accumulation layer that is created at the interface between a TCO layer and an insulator under appropriate electric bias. While the charge density in these accumulation layers has been rigorously determined [2] as ordinarily done in the analysis of MOS capacitors [9], to the best of our knowledge the electromagnetic analysis has been always carried out under the reasonable approximation of a uniform accumulation layer with an equivalent effective permittivity [2, 3], with good agreement with the experimental observations. Nevertheless there are some novel and counterintuitive aspects of the electromagnetic behavior in such systems that are not captured under the uniform layer approximation.

Here a rigorous study of the electromagnetic characteristics of these electron accumulation layers is presented. The unique modal properties of these systems that emerge as a consequence of the graded nature of their permittivity profiles are highlighted. The concept of Accumulation-layer Surface Plasmons is introduced and the conditions for the existence or for the suppression of surface-wave eigenmodes are analyzed.

The system under consideration is shown schematically in figure 1, along with an example of electron density profile and the corresponding electric potential.

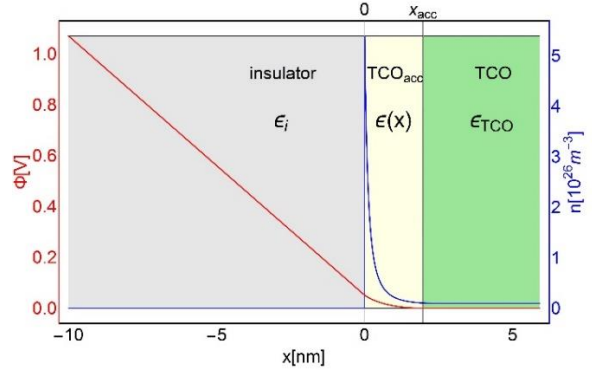

Fig.1 Electrostatic potential and charge density as a function of position in a Silicon-dioxide/Indium Tin oxide MOS structure. The ITO doping concentration is $N_d = 10^{25}[m^{-3}]$ and the applied bias is 1[V].

Following the standard analysis of a MOS capacitor [10], the electron density in the accumulation layer can be expressed as a function of the electric potential $\Phi(x)$ within the TCO as:

$$n(x) = N_d \exp[\Phi(x)/\Phi_T] = N_d \sec^2\left(\frac{x - x_{acc}}{\sqrt{2}L_D}\right) \quad (1)$$

In equation (1) $N_D$ represents the doping density, $\Phi_T = k_B T/q$ is the thermal potential, where $k_B$ is the Boltzmann constant, $T$ is the temperature, and $q$ is the electron charge, $L_D = [\varepsilon_0 \varepsilon_S \Phi_T / (qN_D)]^{1/2}$ is the Debye length, where $\varepsilon_S$ is the static dielectric constant of the TCO, and $x_{acc}$ is the thickness of the accumulation layer given as a function of the surface potential $\Phi_S = \Phi(0)$ by the expression:

$$x_{acc} = \sqrt{2}L_D \arccos\left[\exp(-\Phi_S/\Phi_T)\right] \quad (2)$$

The surface potential $\Phi_S$ appearing in equation (2) is simply related to the applied voltage bias $V$ based on the capacitance of the MOS structure and on the total accumulated charge (obtained integrating equation (1) over the accumulation layer). Under practical experimental conditions the surface potential is of the order of a few $\Phi_T$, leading to an accumulation layer thickness in the range $0.1\,nm < x_{acc} < 2\,nm$ depending on the doping density.

In a time-harmonic regime of the form $e^{-i\omega t}$ the optical characteristics in the TCO region and especially within the accumulation layer are dominated by the free-carriers response, with a relative permittivity well described by the Drude model [1]:

$$\varepsilon(x) = \begin{cases} \varepsilon_{TCO} + \delta\chi \tan^2\left(\dfrac{x - x_{acc}}{\sqrt{2}L_D}\right) & ; \; 0 < x < x_{acc} \\ \varepsilon_{TCO} = \varepsilon_b + \delta\chi & ; \; x > x_{acc} \\ \delta\chi = -\dfrac{\omega_p^2}{\omega(\omega + i\gamma)} \end{cases} \quad (3)$$

In equation (3) $\omega_p = [N_d q^2 / (\varepsilon_0 m^*)]^{1/2}$ is the plasma frequency within the TCO outside of the accumulation layer. Fig.2 shows the permittivity profiles within an ITO layer at a wavelength $\lambda = 1.55\,\mu m$, modeled according to equation (3) with the parameters reported in [1], with three different doping concentrations and biased so as to maintain the surface potential at a value of $\Phi_S = k_B T / q$.

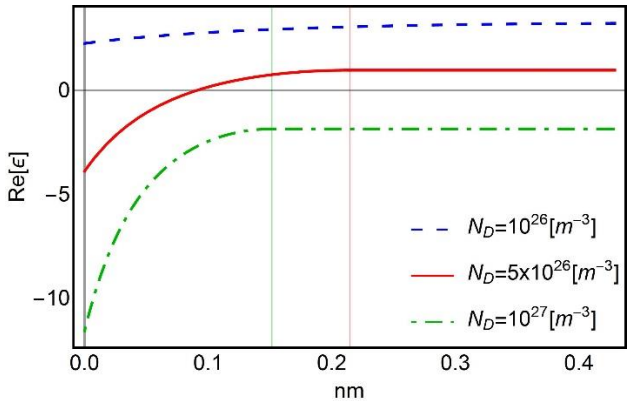

Fig. 2 Permittivity profile within and around the accumulation layer in an ITO film with three different doping concentrations. The surface potential is $\Phi_s = \Phi_T$ in all three cases.

Depending on the doping concentration and on the applied electrical bias it is evident that three qualitatively distinct types of permittivity profile can be obtained at a given frequency:

1. The dashed blue curve in Fig.2 shows that below a certain doping concentration the ITO response is dielectric, with a graded positive permittivity.
2. The dash-dotted green curve in Fig.2 shows that with a sufficiently high doping concentration the ITO response is plasmonic, with a graded permittivity profile that stays negative also outside of the accumulation layer.
3. The red solid curve in Fig.2 shows an intermediate situation in which the ITO response is plasmonic, with a graded negative permittivity, within a portion of the accumulation layer, and dielectric everywhere else, with a graded positive permittivity.

Case 1 is not of particular interest since it is not expected to support any surface eigenmodes (yet it is worth noting that under appropriate electric bias case 1 may be turned into case 3). In the following the cases 2 and 3 will be analyzed.

All the field components associated with a Transverse Magnetic ( TM ) surface wave can be expressed as functions of the magnetic field component parallel to the interface. With reference to the geometry shown in Fig. 1, the magnetic field of a TM surface wave propagating in the $z$ direction has the following form:

$$\mathbf{H}(x,z) = \begin{cases} \hat{\mathbf{y}} H_0 e^{\alpha_i x} e^{i\beta z} & ; \; x < 0 \\ \hat{\mathbf{y}} H_y(x) e^{i\beta z} & ; \; x > 0 \end{cases} \quad (4)$$

In equation (4) $\hat{\mathbf{y}}$ is the unit vector in the $y$ direction, $H_0$ is the complex amplitude of the magnetic field in the insulator right at the interface ($x = 0$), $\beta$ is the propagation constant of the surface mode, $\alpha_i = (\beta^2 - k_0^2 \varepsilon_i)^{1/2}$ is the transverse decay constant in the insulator of permittivity $\varepsilon_i$. Within the inhomogeneous region characterized by the permittivity profile (3) the complex amplitude $H_y(x)$ in the expression (4) satisfies the following equation valid for $x > 0$:

$$\frac{d^2 H_y(x)}{dx^2} - \frac{1}{\varepsilon(x)} \frac{d\varepsilon(x)}{dx} \frac{dH_y(x)}{dx} + [k_0^2 \varepsilon(x) - \beta^2] H_y(x) = 0 \quad (5)$$

The solutions to equation (5) are subject to the following boundary conditions at the interface for $x = 0$:

$$H_y(0) = H_0 \quad ; \quad \left.\frac{dH_y(x)}{dx}\right|_{x=0} = \frac{\alpha_i \varepsilon(0)}{\varepsilon_i} H_0 \quad (6)$$

It is worth noting that due to the nature of the permittivity profile (3), which is such that $\varepsilon(x) \to \varepsilon_{TCO}$ as $x \to x_{acc}$, the solution to the differential equation (5) will have the asymptotic behavior $H_y(x) \sim \exp(-\alpha_{TCO} x)$ as $x \to x_{acc}$, with the decay constant $\alpha_{TCO} = (\beta^2 - k_0^2 \varepsilon_{TCO})^{1/2}$. From this asymptotic behavior the following additional boundary condition can be inferred:

$$\left.\frac{dH_y(x)}{dx}\right|_{x=x_{acc}} = -\frac{\alpha_{TCO} \varepsilon(x_{acc})}{\varepsilon_{TCO}} H_y(x_{acc}) \quad (7)$$

Enforcing the boundary condition (7), instead of solving equation (5) for all $x > 0$, is advantageous for two reasons: 1) it guarantees that the solution is surface-bound (i.e. decaying for $x > x_{acc}$), and

2) it is expedient in view of a numerical solution in that it restricts the range over which equation (5) is to be integrated to $0 < x < x_{acc}$.

In case 2 ( $\varepsilon(x) < 0$ for every $x > 0$ ), equation (5) does not have any singular points and it may be integrated numerically using for instance the Runge-Kutta method [11] for the initial value problem constituted by equation (5) and (6), and the shooting method [12] to find the eigenvalues $\beta$ complying with the condition (7). While a fully numerical approach can provide an accurate description of the electromagnetic eigenmodes of the accumulation layer for specific sets of parameters, obtaining analytical approximate solutions could shed light on the physics of the problem and give a sense of the functional dependence on the relevant parameters (such as doping density, electric bias, and superstrate permittivity).

Equation (5) cannot be directly integrated in closed form, but it lends itself to a perturbative approximation that greatly simplifies the analytical treatment of the problem. The key observation is that the accumulation layer width $x_{acc}$ is much smaller than the wavelength $\lambda$. As a consequence the spatial variations of the magnetic field over the accumulation layer are expected to be dictated by the fast spatial variation of the permittivity profile $\varepsilon(x)$ rather than by wave dynamics represented by the third term in equation (5). Neglecting such term in equation (5) the following approximate equation for the magnetic field within the accumulation layer is obtained:

$$\frac{d^2 H_y(x)}{dx^2} - \frac{1}{\varepsilon(x)} \frac{d\varepsilon(x)}{dx} \frac{dH_y(x)}{dx} = 0 \quad (8)$$

Equation (8), along with the boundary conditions (6) at $x = 0$ can be integrated in closed form to yield the following analytical solution:

$$H_y(x) = H_0 + H_0 \frac{\sqrt{\beta^2 - k_0^2 \varepsilon_i}}{\varepsilon_i} (\varepsilon_{TCO} - \delta\chi) x +$$
$$+ H_0 \frac{\sqrt{\beta^2 - k_0^2 \varepsilon_i}}{\varepsilon_i} \sqrt{2} L_d \delta\chi \left[ \tan\left(\frac{x_{acc}}{\sqrt{2}L_d}\right) + \tan\left(\frac{x - x_{acc}}{\sqrt{2}L_d}\right) \right] \quad (9)$$

The expression (9) is valid within the accumulation layer, i.e. for $0 < x < x_{acc}$. Beyond the accumulation layer, for $x > x_{acc}$, the permittivity takes the constant value $\varepsilon_{TCO}$ as indicated in formula (3) and the field assumes the common decaying exponential form $H_y(x) = H_y(x_{acc}) \exp[-\alpha_{TCO}(x - x_{acc})]$. The eigenvalue $\beta$ can be determined by imposing the boundary condition (7):

$$H_0 \frac{\sqrt{\beta^2 - k_0^2 \varepsilon_i}}{\varepsilon_i} \varepsilon_{TCO} = -\sqrt{\beta^2 - k_0^2 \varepsilon_{TCO}} H_y(x_{acc}) \quad (10)$$

Equation (10) can be solved for $\beta$ by obtaining from equation (9) the value $H_y(x_{acc})$ of the magnetic field at the edge of the accumulation layer. The expression of $\beta$ that can be so obtained is in closed form, but extremely cumbersome. On the other hand solving equation (10) through a perturbation series in the parameter $x_{acc}/\lambda$ up to the first order leads to a simpler and physically more transparent solution that is still extremely accurate at wavelengths away from the resonance condition $\varepsilon_{TCO} + \varepsilon_i = 0$:

$$\beta = \beta_0 + \frac{\beta_0^3 \left[ x_{acc} \varepsilon_b + \sqrt{2} L_d \delta\chi \tan\left(\frac{x_{acc}}{\sqrt{2}L_d}\right) \right]}{k_0 \sqrt{(\varepsilon_{TCO}^2 - \varepsilon_i^2)(\varepsilon_i - \varepsilon_{TCO})}} \quad (11)$$

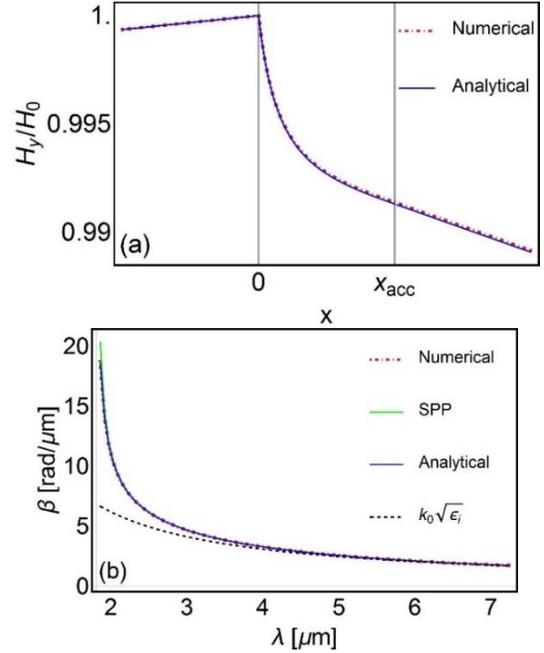

Fig.3 (a) Magnetic field profile and (b) propagation constant of an ASP in a SiO$_2$-ITO MOS structure with doping concentration $N_d = 10^{27} [m^{-3}]$ and surface potential $\Phi_s = 3\Phi_T$.

In equation (11) the unperturbed solution $\beta_0 = k_0 [\varepsilon_i \varepsilon_{TCO} / (\varepsilon_i + \varepsilon_{TCO})]^{1/2}$ is the dispersion relation at the interface between the dielectric $\varepsilon_i$ and a plasmonic medium with uniform relative permittivity $\varepsilon_{TCO}$. The accuracy of the results (9) for the fields and (11) for the dispersion are verified in Fig. 3(a) and (b) against the shooting-method/Runge-Kutta numerical solution of the exact equation(5). As evident from Fig.3, the agreement between the numerical and the analytical solution is excellent.

Based on the nearly perfect agreement with the numerical solution of the exact equations, the expression (9) is a faithful representation of the class of surface eigenmodes supported by an accumulation layer, or Accumulation-layer Surface Plasmons (ASPs). It is interesting to notice that while the field profile of an ASP within the accumulation layer is markedly different from the exponentially decaying field of a surface plasmon polariton (SPP), the dispersion relation (11) is to an excellent approximation identical to the one of the SPP supported in the absence of the accumulation layer (i.e. without electrical bias). That is no longer true when the resonance condition $\varepsilon_{TCO} + \varepsilon_i = 0$ is approached, in which case, due the strong localization of the fields in proximity of the interface, the carrier density in the accumulation layer has a pronounced effect on the ASP dispersion.

The solid curves in Fig. 4 show the normalized propagation constant and the normalized propagation length of an ASP near resonance as a function of the surface potential obtained by solving the exact equation(10). By comparison the long-dashed curves show that, as expected, the perturbative approximation (11) is no longer accurate near resonance. The dotted curves in Fig. 4 correspond to the propagation of a SPP in the absence of the accumulation layer. The large excursion in the ASP propagation constant as a function of the applied electric bias makes these surface-wave modes interesting candidates for the realization of ultra-compact phase-modulators.

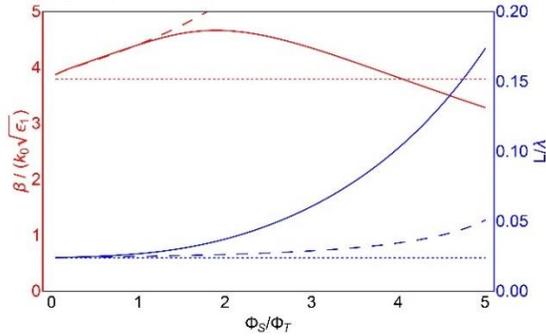

Fig.4 Normalized propagation constant and propagation length of an ASP in a SiO$_2$-ITO MOS structure with doping concentration $N_d = 10^{27} [m^{-3}]$ at the near resonant wavelength $\lambda = 1.8 [\mu m]$.

Care must be taken when analyzing the modal properties of the system in case 3, in which the real part of the permittivity is negative within a portion of the accumulation layer and smoothly transitions to a constant value $\varepsilon_{TCO} > 0$ beyond of the accumulation layer for $x > x_{acc}$. This is a rather unique situation, as within the same medium three distinct electromagnetic responses coexist: plasmonic near the interface with the insulator, dielectric within the TCO bulk, and epsilon-near-zero (ENZ) [13] in between. In this instance, approximating the accumulation layer by a uniform medium with an average permittivity leads to conclusions that are erroneous, not only quantitatively, but also qualitatively. A uniform plasmonic thin film of thickness $x_{acc}$ and negative permittivity $\varepsilon_p$ sandwiched between two dissimilar dielectric with positive permittivities $\varepsilon_i$ and $\varepsilon_{TCO}$, such that $|\varepsilon_p| > \max(\varepsilon_i, \varepsilon_{TCO})$ will support at least one mode [14], with the remarkable property of displaying more confinement (i.e. higher $\beta$) as the film thickness is reduced. As shown in the following such mode is not supported by the accumulation layer, and the reason may be identified in the presence of an ENZ region.

Attempting a shooting-method/Runge-Kutta numerical solution equation (5) does not yield any surface-bound eigenmodes. In order to elucidate the mechanism suppressing ASPs, equation (5) is considered under the following simplifying assumptions: the permittivity in the accumulation layer is assumed to be real and the permittivity profile is linearized around the point $x_0$ where $\varepsilon(x_0) = 0$. Under these approximations equation(5) has a regular singular point at $x_0$, so a solution can be obtained in the form of a Frobenius series[15] $H_y(x) = (x - x_0)^\gamma \sum_{n=1}^{\infty} A_n (x - x_0)^n$. The parameter $\gamma$ in this equation is one of the two roots of the indicial polynomial $P(\gamma)$ associated to the Frobenius series solution of (5). In order to fulfill the boundary conditions (6) and (7) a second linearly independent solution associated with the second root of the indicial polynomial must be obtained. Under the present assumptions $P(\gamma)$ has two identical roots, so it can be shown that the second solution has a logarithmic singularity at $x_0$. Because of its singular nature such second solution must be discarded as unphysical, and therefore no guided eigenmodes exist in this case.

In conclusion a rigorous analytical study of the eigenmodes supported by a charge accumulation layer was presented. The new class of surface plasmons termed Accumulation-Layer Surface Plasmon (ASP) was introduced. Near resonance ASP display a vast effective index tunability that could be employed for the realization of ultra-compact phase modulators. The suppression of ASP in the presence of ENZ region was also discussed. The present analytical results provide important guidelines for the design of TCO-based optoelectronics devices – a topic of intense current interest.

**Funding.** Airforce Office of Scientific Research (AFOSR) (FA9550-16-1-0152)

## Extended References.